\documentclass[%
 amsmath,amssymb,
reprint, 
author-numerical,%
superscriptaddress,
]{revtex4-2}

\usepackage{graphicx}
\usepackage{dcolumn}
\usepackage{bm}
\usepackage{color}

\begin{document}


\title{A Local-Phase Framework for the BaTi$_{1-x}$Zr$_x$O$_3$ phase diagram: From Ferroelectricity to Dipolar Glass
}


\author{M. Sepliarsky\thanks{corresponding author}}
\email{sepliarsky@ifir-conicet.gov.ar}
\affiliation{Instituto de Física Rosario (CONICET-UNR) and Facultad de Ciencias Exactas, Ingenier\'\i{}a y Agrimensura,Universidad Nacional de Rosario, Rosario, Argentina}
\author{F. Aquistapace}%
\affiliation{Department of Physics and Astronomy “Galileo Galilei”, University of Padova, Padova, Italy}
\author{F. Di Rino}
\affiliation{Institute of Physics of the Czech Academy of Sciences, Na Slovance 2, Prague 8, 182 00, Czech Republic}
\author{R. Machado}%
\affiliation{Instituto de Física Rosario (CONICET-UNR) and Facultad de Ciencias Exactas, Ingenier\'\i{}a y Agrimensura,Universidad Nacional de Rosario, Rosario, Argentina}
\author{M. G. Stachiotti}
\affiliation{Instituto de Física Rosario (CONICET-UNR) and Facultad de Ciencias Exactas, Ingenier\'\i{}a y Agrimensura,Universidad Nacional de Rosario, Rosario, Argentina}

\date{\today}

\begin{abstract}

We apply a first-principles-based atomistic model to investigate the BaTi$_{1-x}Zr_xO_3$ phase diagram, focusing on both macroscopic and local structural changes. Our approach, which combines molecular dynamics with machine learning techniques, accurately captures the influence of Ti and Zr cations on their local environment and its evolution with composition and temperature. The computed phase diagram shows excellent agreement with existing experimental and theoretical data. Beyond reproducing known results, our analysis reveals that the behavior of the solid solution across different compositions and temperatures can be understood in terms of coexisting Ti cells with different symmetries, whose stability depends on the local B-site configuration.
This local-phase-based approach provides a unified description of the distinct regions of the solid solution—including ferroelectric, relaxor, and dipolar glass phases—and captures the continuous evolution from one regime to another. Our findings demonstrate how atomic-level distortions drive the complex macroscopic behavior of the  material.


\end{abstract}

\maketitle


\section{\label{intro}Introduction}

BaTiO$_3$-based materials have attracted considerable attention due to their outstanding ferroelectric and dielectric properties, making them highly suitable for a wide range of applications, including capacitors, sensors, actuators, electro-optic devices, and energy harvesting systems\cite{acosta_17,rodel_09}. BaTiO$_3$ (BTO) is a prototypical perovskite oxide that undergoes a series of temperature-dependent phase transitions, from a high-temperature cubic paraelectric phase (C) to ferroelectric tetragonal (T), orthorhombic (O), and rhombohedral (R) phases upon cooling \cite{kwei_93}. 
However, pure BTO exhibits intrinsic limitations that can restrict its utility in advanced devices. To address these drawbacks and tailor its functional properties, BTO is often chemically modified, either through doping or by forming solid solutions with other perovskite oxides\cite{veera_20}. 
Among them,  Ba(Zr$_x$Ti$_{1-x}$)O$_3$ (BZT), the solid solution of BTO and the nonpolar BaZrO$_3$ (BZO),
has attracted considerable attention due to its tunable dielectric and ferroelectric properties, which make it a promising lead-free alternative for a variety of applications. Its ability to exhibit both classical ferroelectric and relaxor behavior, depending on composition and temperature, allows for tailored responses suitable for specific applications. 
It is also recognized as a model system for investigating the transition from ferroelectric to relaxor behavior via isovalent substitution, in the absence of nominal charge disorder. 

The phase diagram of BZT, as inferred from dielectric measurements, reveals a compositional evolution that can be divided into distinct regimes depending on the Zr content\cite{petz_21,maiti_11,dong_12,shva_09}. At low concentrations ($x<0.11$), the solid solution behaves as a typical ferroelectric, similar to BTO, with well-defined phase transitions. In the intermediate range ($0.11\leq x\leq 0.27$), the system exhibits a diffuse ferroelectric transition, marked by a broadening of the dielectric permittivity peak with temperature. For $x >0.27$, BZT displays characteristic relaxor behavior: the permittivity maximum ($T_m$) becomes frequency-dependent, shifting to higher temperatures with increasing frequency, while the macroscopic structure remains cubic throughout the temperature range. At even higher Zr concentrations ($x> 0.8$), the system enters a dipolar glass state and eventually behaves as a normal dielectric at $x=1$.


While the macroscopic behavior of BZT is well-mapped—with distinct ferroelectric, relaxor, and dipolar glass regimes—the underlying local structural origins remain a subject of debate. 
At the core of this controversy lies the precise nature of the polar evolution from the ferroelectric state to the relaxor regime. In this transition, local dipoles that are stabilized by long-range order are often described as transforming into disordered polar nanoregions (PNRs) within an overall nonpolar cubic symmetry.
Evidence from local probes, such as Raman spectroscopy and XAFS, has provided valuable but seemingly contradictory insights. Some studies strongly support a rhombohedral local distortion \cite{dix_06,busca_14}, arguing that the Ti ions are displaced along the ⟨111⟩ axes, akin to the low-temperature phase of pure BTO. This view posits that the BZT relaxor phase consists of small, frozen PNRs with rhombohedral symmetry \cite{akba_12}. The randomness introduced by Zr substitution is thought to prevent these regions from coherently aligning to form a long-range ferroelectric domain, leading to the characteristic relaxor behavior.
However, more recent studies, combining evidence from local structure probes and first-principles calculations, have revealed a more complex scenario. They show that the local behavior arises from the competing effects of Ti and Zr ions at the B-site \cite{lau_09,lau_2011,levin_11,amor_18,lau_10,jeong_10}, even though both cations are homovalent. First, their distinct electronic configurations lead to markedly different local structural behaviors: Ti's tendency to hybridize with oxygen favors off-center displacements, while Zr ions tend to remain centered. Second, the difference in their ionic radii creates local strain effects. In addition to these intrinsic differences, the spatial arrangement of B-site cations further influences the solid solution's behavior. For instance, isolated Ti atoms relax to the centrosymmetric position, yet near-neighbor pairs exhibit parallel ⟨001⟩ displacements.  The direction of a Ti displacement is strongly influenced by the Ti/Zr distribution in adjacent unit cells, giving evidence for four types of polar displacements: along directions  close to the 〈111〉, 〈011〉, or 〈001〉 crystallographic axes, or almost canceled  \cite{lau_10}.  These local structural variations suggest the coexistence of distinct polar and non-polar regions, pointing to a complex nanoscale phase coexistence characteristic of relaxor behavior. In fact, the diffuse phase transitions observed in BZT ceramics are shown to be caused by coexisting ferroelectric and paraelectric phases, which can be described by a normal distribution of Curie temperatures \cite{hen_1982}.

Given this complexity and the observed link between local structural distortions and macroscopic properties, an atomic-level perspective is essential. While the behavior of BZT at finite temperatures has been theoretically investigated using methods like phase-field models \cite{hua_20} and the effective Hamiltonian method \cite{akba_05,ment_19,mayer_22,lader_23}, these approaches often introduce additional terms or concentration-dependent parameters to account for local effects. In the present work, we adopt a different strategy by explicitly treating the individual atoms and their interactions. We introduce a first-principles-based atomistic framework to investigate the BZT phase diagram, explicitly identifying and characterizing distinct local structural phases as a function of composition and temperature by combining molecular dynamics and machine learning (ML) techniques. 
Over recent years, several machine learning (ML) algorithms have been developed for atomic structure identification\cite{schmi_19,zilet_18,elap_22}. Among them, clustering techniques are particularly useful to identify patterns and to group data points that share similar features\cite{ester_96,frey_07}. Here, we use MultiSOM\cite{aquis_23,aquis_24} for a systematic identification of distinct polar or structural phases in BZT. This approach bridges the gap between atomic-level disorder and the macroscopic behavior, providing a unified local-phase-based framework that captures the continuous evolution from ferroelectric to relaxor and dipolar glass behavior.

\section{\label{sec:model}Model and computational details}

We use shell model potentials  that  were developed to study the behavior of both end-member compounds,  BTO ~\cite{macha_19} and BZO \cite{sepli_23}. In the description, each atom is modeled  as a pair of charged particles: a core connected to a massless shell, which phenomenologically represents atomic polarizability. The model also considers electrostatic interactions between the cores and shells of different atoms, as well as short-range interactions between the shells. The parameters for the models were fitted to first-principles results on key properties of each compound under the assumption of transferable interactions. Since both compounds share the same ABO$_3$ perovskite structure and differ only in the element at the B-site, the interactions common to both (O-O and Ba-O) are the same, while those involving the B-cation differ. The approach assumes that the interaction between atoms does not strongly depend on their local environment and remains valid in conditions beyond those considered during the fitting process. This makes it possible to simulate the Ba(Zr$_{x}$Ti$_{1-x}$)O$_3$ solid solution over the whole concentration range without the need for extra parameters.
Additional details regarding the development and validation of the model can be found in Refs. ~\cite{macha_19} and \cite{sepli_23}, where the full set of parameters is also reported.

We apply the interatomic potentials to simulate the finite temperature behavior of BZT as a function of the composition. Molecular dynamics (MD) simulations are carried out using the DL-POLY code~\cite{dlpoly} within a constant stress and temperature (N,$\sigma$,T) ensemble. In this way, the size and shape of the simulation cell are dynamically adjusted in order to obtain the desired average pressure. A supercell of $20\times 20\times20$ 5-atom unit cells (40000 atoms) is used with periodic boundary conditions. The runs are made at temperature intervals of 10~K, and with a time step of 0.4~fs. Each MD run consists of at least 60000 time steps for data collection after 20000 time steps for thermalization. We estimate the most stable states as zero-temperature limit MD simulations. To avoid high-energy metastable states, heating and quenching were repeated until the forces acting on individual ions were less than 0.01 eV \AA$^{-1}$.

To characterize the local behavior of the system, we define the local polarization as the dipole moment per unit volume of a perovskite cell considered here as centered at the B-site (Ti or Zr atoms) and delimited by the 8 Ba near neighbors at the corners of the box.
In the calculations, we take the contributions from all atoms in the conventional cell and the atomic positions with respect to this center\cite{sepli_11}: 
\begin{equation}
\vec{p}=\frac{1}{v}\sum \frac{z_i}{\omega _i}\left ( \vec{r}_i-\vec{r}_{\rm{B}} \right )
\label{SMpol}
\end{equation}
\noindent where $v$ is the volume of the cell, $z_i$ and $\vec{r}_i$ denote the charge and the position of the $i$-th particle, respectively, $\vec{r}_{\rm{B}}$ denote the B-type atom position, and ${\omega _i}$ is a weight factor equal to the number of cells to which the particle belongs.
In this work, we consider only BZT configurations with random B-site occupancy and do not explicitly introduce regions enriched in either Ti or Zr.  Nonetheless, local compositional fluctuations can still emerge spontaneously due to statistical variations. 
At each concentration, we examine three configurations with different random distributions of Ti and Zr ions. In addition, two initial atomic arrangements are considered: one with atoms placed at their ideal cubic positions and another with Ti ions initially displaced off-center along the [111] direction to favor polar order. 

\section{\label{sec:results} Results }
\subsection{\label{sec:p0k} Zero-temperature properties}
 
\subsubsection{Macroscopic properties}

\begin{figure}
\centering
\includegraphics[width=\linewidth]{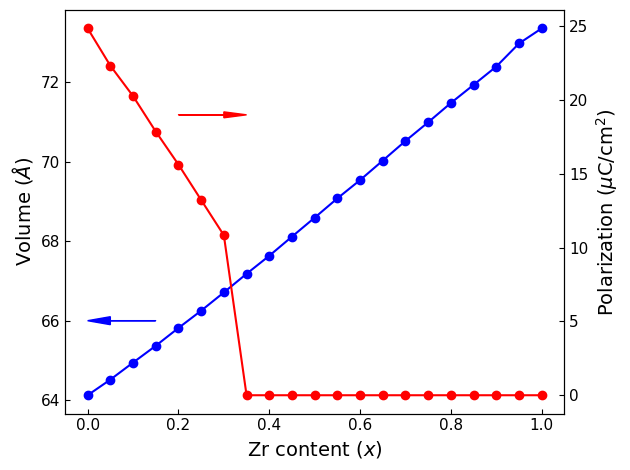}
\caption{\label{fig:Vol}Simulated average cell volume and net polarization of BaZr$_x$Ti$_{1-x}$O$_3$  as a function of composition. }
\end{figure}

Fig. \ref{fig:Vol} shows the model results for the composition dependence of the unit cell volume and the net  polarization in BZT at T = 0 K. Since the resulting values are not affected by the specific B-site cation arrangement, only one representative point is shown at each composition. The volume exhibits a clear linear dependence on the Zr content, following Vegard's law in excellent agreement with experiments. The simulated values range from 64.0 \AA$^3$ for BTO to 73.35 \AA$^3$ for BZO, closely matching the experimental range of 64.0 \AA$^3$ to 73.60 \AA$^3$ ~\cite{akba_05}.
This composition-driven volume variation induced by the size mismatch between Zr and Ti cations constitutes a key factor in understanding the origin of relaxor behavior in the systems. The ability of the model to reproduce this trend provides a solid basis for investigating the local structural effects that govern the observed macroscopic response.

The figure also shows the net polarization as a function of the composition, revealing two distinct regions.  At low Zr concentrations $x \le 0.3$, the system exhibits a ferroelectric phase with polarization along the [111] direction, consistent with pure BTO. Above $x = 0.3$, the cubic, nonpolar structure becomes the energetically favorable one. This transition point agrees well with the experimental crossover at $x = 0.27$ obtained from dielectric measurements\cite{simon_04}. This close agreement demonstrates the capability of the model to accurately reproduce the ferroelectric-to-relaxor transition  characteristic of BZT with increasing Zr content. 
It is worth noting that near the transition, polar and non-polar configurations become nearly degenerate in energy, with stable instances of both phases obtained on either side of the crossover. Such behavior reveals a gradual transition characterized by phase competition, where the polar configuration remains slightly energetically favored up to $x = 0.3$.

\subsubsection{Local Structure Analysis}
The macroscopic behavior described above originates from the effects that take place at the local level. To uncover the microscopic mechanisms behind it, we begin by analyzing the pair distribution function (PDF). This quantity allows us to identify subtle changes in short-range order induced by the presence of Ti and Zr cations.
\begin{figure}
    \centering
    \includegraphics[width=1.0\linewidth]{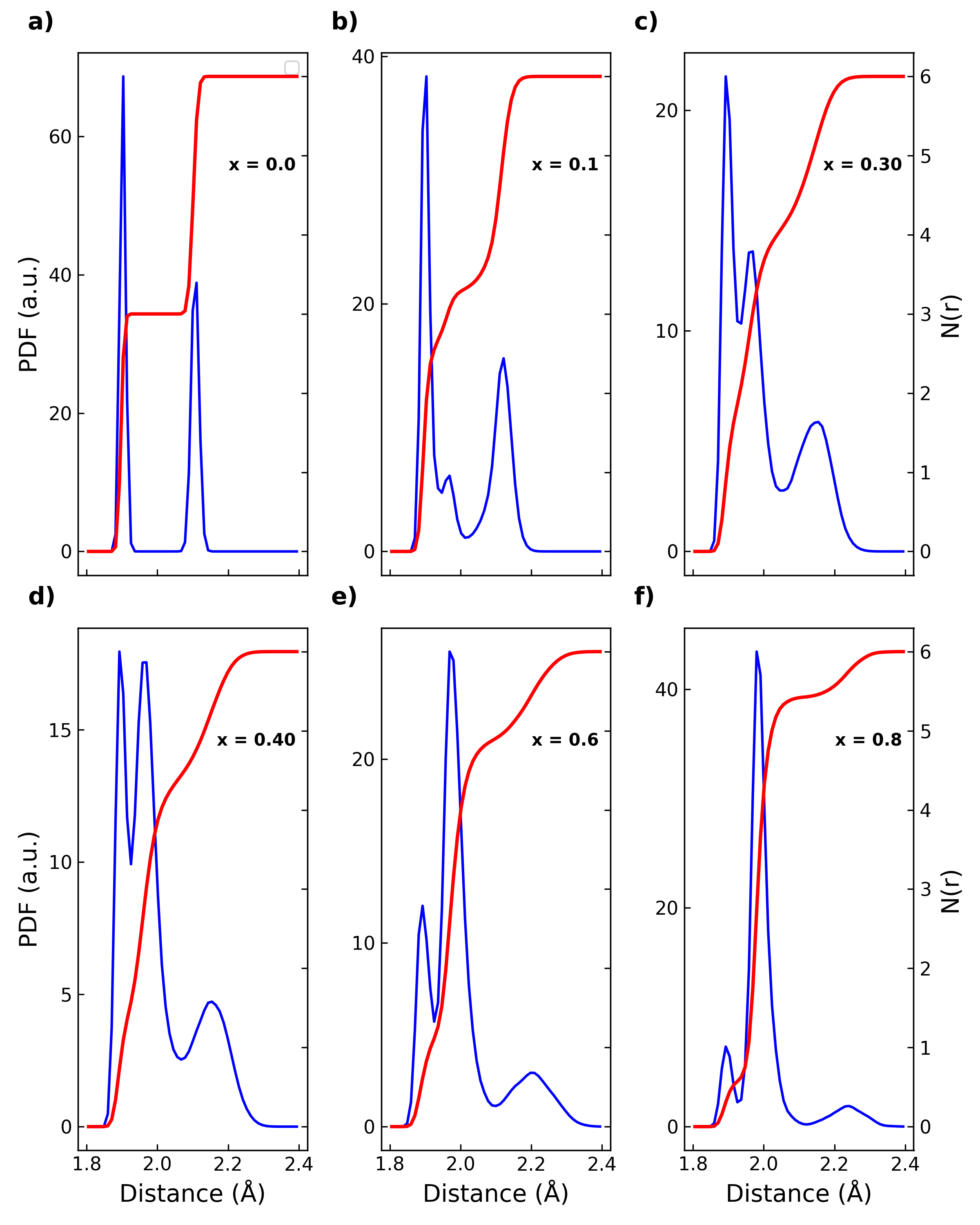}
    \caption{Pair distribution function of Ti–O pairs (blue) and corresponding integrated coordination number (red) for different Zr concentrations.}
    \label{fig:RDF-TiO}
\end{figure}
\begin{figure}
\centering
\includegraphics[width=\linewidth]{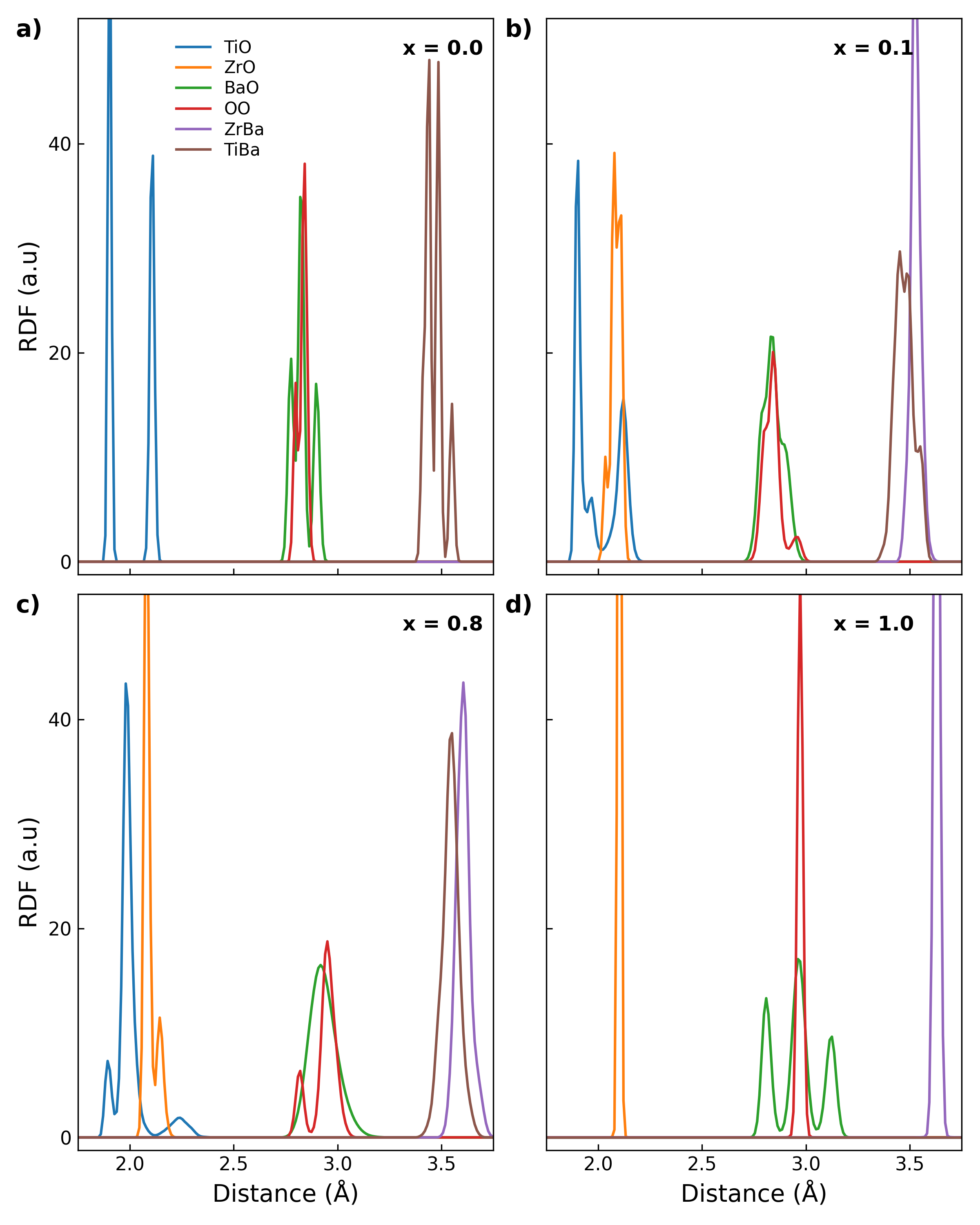}
\caption{\label{fig:RDF}Pair distribution function (PDF) obtained from the simulation with the atomistic model for BZT at $x=$ 0.0, 0.1, 0.8 and 1.0}
\end{figure}

Fig. \ref{fig:RDF-TiO} displays the PDF for the Ti-O pair (in blue) and its integrated values (in red) for different Zr concentrations. The curves are presented at a low temperature of 5 K rather than the zero-temperature limit. This approach allows the minimal thermal fluctuations to broaden the peaks, providing a clearer and more realistic view of the interatomic distances compared to the sharp, idealized delta functions that would be present at T = 0 K.

In pure BTO ($x$=0), the off-centering of Ti atoms within the oxygen octahedra produces distinct short and long Ti–O bonds, which manifest as the characteristic double peak of the R phase.
Each peak has an integrated value of 3. As the Zr concentration increases, the peak associated with long bonds gradually decreases in intensity, while an additional peak emerges at an intermediate distance. Although it is not possible to definitively assign the PDF curves to a single phase, the appearance of this intermediate peak is consistent with the development of local distortions compatible with O, T, or even C-like symmetries. For instance, the PDF for $x$ =0.3 shows a predominantly O-like structure, displaying three peaks, each with an integrated value close to 2. The local structure evolves as the Zr concentration is increased. The intermediate concentration of $x$=0.4 indicates a mixture of O and T character. At $x$=0.6, the PDF curve displays a predominantly T-like character with three peaks and integrated values close to 1, 4, and 1, respectively. At $x$=0.8, the predominant local structure still has tetragonal symmetry, but the integrated value of the intermediate peak is larger than 4, indicating a mixture with the non-polar cubic phase.

Another key feature evident in Fig. \ref{fig:RDF-TiO} is the expansion of the Ti-O bond length distribution with increasing Zr content. This broadening arises from the strong Zr-O repulsion, which compresses the adjacent Ti-O bond along the Ti-O-Zr axis. This axial compression, in turn, promotes a lateral expansion in the perpendicular plane, effectively creating more space for Ti to shift off-center transversely. These two coupled effects, the inhibition of Ti displacements along the Ti-O-Zr bond and their enhancement in the perpendicular direction, highlight the anisotropic influence of Zr on the local structural environment.

The distributions for the other atom pairs are presented in Fig. \ref{fig:RDF}, which displays the PDF for the end-members and two intermediate compositions. For completeness, the Ti–O pair is also included. The distributions for the Zr-O pair (orange) at $x$=0.10 and $x$=0.80 closely resemble those of pure BZO. However, both compositions show a broader distribution with a shoulder, indicating slight displacements of the Zr atoms. These distortions, which arise from the local presence of Ti, are likely responsible for the induced polarizations in Zr cells. 
Notably, the average Ti–O and Zr–O bond lengths are close to those of the pure compounds. Thus, the interatomic distances between cations and their nearest-neighbor oxygens remain nearly unchanged in the solid solution compared to the end members, BTO and BZO, consistent with EXAFS measurements\cite{lau_2011}.
However, a slight shift to longer distances is observed in the Ti–O pairs with increasing Zr content, while a shift to shorter distances appears in the Zr–O pairs at low Zr concentrations. 
These subtle variations reflect local structural adjustments and highlight the asymmetric influence of the surrounding cation environment on bond lengths.
For the O–O pairs, each pure compound exhibits a single, well-defined peak corresponding to the characteristic oxygen–oxygen distance within the octahedra. In intermediate compositions, the pair distribution shows a superposition of two peaks, with intensities that scale proportionally with the concentration. This behavior reveals the presence of two distinct types of octahedral environments: one centered on Ti and the other on Zr.  The difference between the peaks arises because oxygen atoms belonging to Ti-centered octahedra are slightly closer to each other than those in Zr-centered ones. As a result, the oxygen sublattice exhibits a mixed character across the solid solution, combining structural features of both end members.

The Ba–O pair distribution broadens toward longer distances with increasing Zr content, a result of the expanded local environment around Ba due to the incorporation of larger Zr ions. In BZO, the splitting into three distinct peaks corresponds to correlated rotations of the oxygen octahedra, as predicted by our model\cite{sepli_23}. Although the stability of these distortions remains a topic of debate, they become indistinguishable upon the introduction of small amounts of Ti. Consequently,  we consider that they do not significantly influence the overall structural behavior observed in the solid solution.
Finally, the Ba–Ti and Ba–Zr pair distributions closely resemble those of the pure compounds, suggesting that Ba atoms can easily adapt to their local environments. This structural flexibility of the A-site cation helps maintain nearly constant Ba–B (B = Ti, Zr) distances throughout the entire composition range. 

\begin{figure}
\centering
\includegraphics[width=\linewidth]{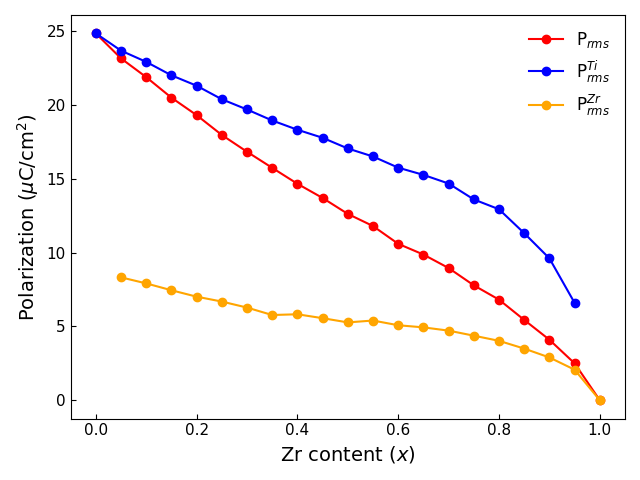}
\caption{\label{fig:P0}Root mean square of local polarization as a function of Zr concentration at T = 0 K for the complete system (red), Ti cells (blue), and Zr cells (yellow). }
\end{figure}

The gradual transformation of the local structure with composition is also evident when examining the local polarization. In particular, the root mean square of the local (per-cell) polarization, $P_{rms}$, provides a measure of the magnitude of local dipoles regardless of their orientation. As shown in Fig. \ref{fig:P0}, $P_{rms}$ decreases progressively with increasing $x$, yet remains finite even within the non-polar regime. 
This smooth decline, rather than an abrupt vanishing, highlights the role of local structural effects and the growing influence of polar disorder, both of which align with the onset of relaxor-like behavior. 
A similar trend emerges when the two types of B-site cells are analyzed separately. Ti-centered cells, which dominate the polar response, maintain large polarization values even at high $x$, revealing their strong intrinsic tendency to polarize. In contrast, Zr-centered cells exhibit much smaller polarization, arising mainly from distortions induced by neighboring Ti-rich regions.

\subsubsection{\label{sec:ML} Clustering analysis of local polarization patterns}

To gain deeper insight into the microscopic characteristics of the system and quantify the coexistence of phases, we employed the unsupervised machine learning method MultiSOM ~\cite{aquis_23}  to classify individual cells into distinct local phase types based on their polarization vectors. We leveraged the Ovito software\cite{ovito} to calculate the necessary descriptors and to analyze the resulting classifications.

We focus the analysis exclusively on Ti cells, which are polar active ones. Zr cells are excluded due to their intrinsically nonpolar nature, which results in an inconsistent or ambiguous classification.
MultiSOM, like other clustering algorithms, aims to provide an optimal representation of the underlying sample space. Here, we consider a set of observations (Ti-centered cells) that are mapped onto a one-dimensional space defined by the magnitude of their polarization $p_i$. We enforce the algorithm to partition this space into two groups: polar and nonpolar cells. 
It should be emphasized that the classification into “polar” and “non-polar” cells is the outcome of an unsupervised clustering procedure and therefore reflects a statistical partition of the data rather than a sharp physical distinction. 
The data-driven polarization threshold  $p=$13 $\mu C/cm^{2}$ arises as the equidistant point between the two centroids determined by the algorithm, and its value does not necessarily correspond to a physically meaningful threshold. While cells with $p>$13 $\mu C/cm^{2}$
are more likely to exhibit robust dipoles, this criterion cannot unambiguously distinguish between truly ferroelectric cells and those with smaller polarization that may result from local fluctuations or induced effects. Thus, the “polar/non-polar” terminology should be understood as a convenient classification imposed by the algorithm, rather than a strict physical definition.
Next, the code enables the separation of the three polar phases (T, O, and R) by using the dipole components as descriptors in the following way:$(p_{max}, p_{min}, p_{mid})$. Here, $p_{max}$ denotes the component with the largest absolute value, $p_{min}$ the one with the smallest absolute value, and $p_{mid}$ the remaining component. All values are normalized with respect to $p_{max}$. This representation of the polar vector facilitates the classification since it is symmetric under rotations, and it effectively reduces the dimensionality of the representation space to two variables, $(p_{min}, p_{mid})$, as the normalized $p_{max}$ is always equal to 1. 
The classification can be understood as a tessellation of this two-dimensional space into three distinct regions. The region corresponding to the T phase is approximately defined by $p_{mid} < 0.5$ and $p_{min}<0.35$. The O phase is characterized by $0.5<p_{mid} \leq 1$ and $p_{min} <0.35$. Finally, the R phase is identified approximately by $0.35 < p_{mid} \leq 1$ and $0.35< p_{min} \leq 1$. As with the distinction between polar and non-polar phases, the assignment of T, O, and R phases should also be regarded as an interpretation of the groups identified by MultiSOM, rather than a strict definition.

\begin{figure}
\centering
\includegraphics[width=\linewidth]{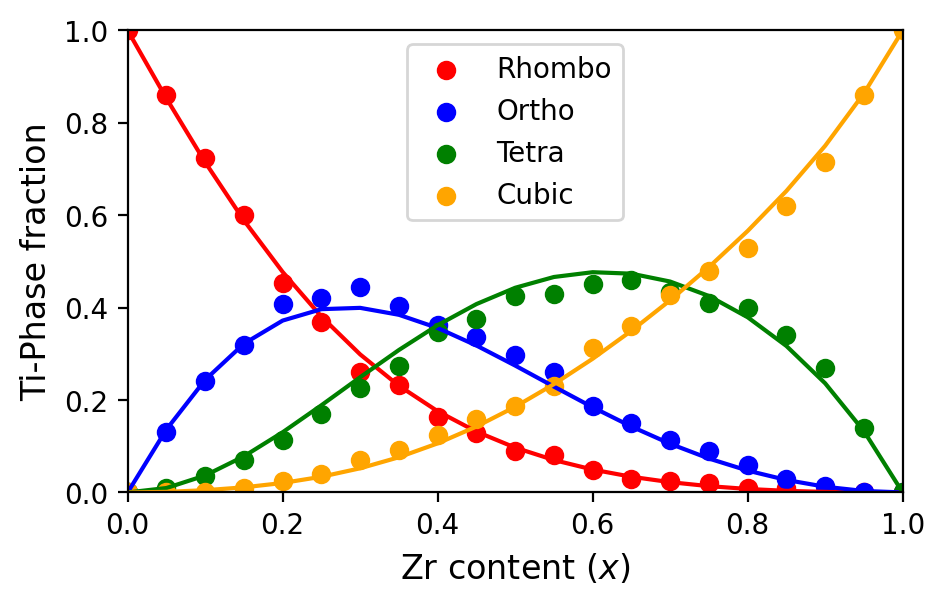}
\caption{\label{fig:Ph_x} Phase fractions of Ti-centered unit cells in BaZr$_x$Ti$_{1-x}$O$_3$ as a function of Zr composition at $T = 0$ K. Symbols correspond to the clustering analysis, while solid curves represent the fit from the probabilistic model. }
\end{figure}

Fig. \ref{fig:Ph_x} displays the evolution of the fraction of each phase among Ti cells as a function of composition at T = 0 K. Symbols correspond to the clustering analysis, while solid lines represent the fit of the probabilistic model. The results reveal that the various ferroelectric phases found in pure BTO coexist across a broad composition range in BZT. Their relative abundance evolves continuously with composition, reflecting the progressive transformation of the local structure. As the Zr content increases, the dominant local symmetry shifts successively from R to O, then to T, and finally to C, resembling the sequence of phase transitions observed in BTO with increasing temperature.

The distribution of these phases at each concentration can be rationalized by assuming that the local polarizations (or equivalently, the displacements of Ti atoms) are entirely determined by the Ti/Zr distribution in the adjacent unit cells\cite{lau_10,lau_2011}. 
For this purpose we model the system as a simple cubic lattice in which each site is occupied by either a Ti or a Zr ion, and each Ti interacts only with its six nearest neighbors (along the $\pm x$, $\pm y$, and $\pm z$ directions). Within this framework, Ti ions are treated as active sites capable of sustaining local polarization, whereas Zr ions act as inactive sites.
For a given concentration $x$ with a random distribution of Ti and Zr, the probability that a site has exactly \( k \) Ti neighbors (out of 6 possible) follows a binomial distribution:

\begin{equation*}
p_x(k) = \binom{6}{k} \, x^k \, (1 - x)^{6 - k}
\end{equation*}

\noindent where $k \in \{0, 1, \ldots, 6\}$. Then, the overall probability of observing a given local phase \( \mathcal{P} \) at a given concentration \( x \) is:

\begin{equation*}
P_x(\mathcal{P}) =  \sum_{k=0}^6 A_\mathcal{P}(k)\, p_x(k) = \sum_{k=0}^6 \binom{6}{k} A_\mathcal{P}(k)\, x^k (1 - x)^{6 - k}
\end{equation*}

\noindent where $A_\mathcal{P}(k)$ is the conditional probability that a site exhibits a phase $\mathcal{P}$, given that it has $k$ Ti neighbors. 
These coefficients establish the connection between the local chemical environment and the resulting structural phase. It can be noticed that this formulation imposes a constraint on the coefficients $A_\mathcal{P}(k)$, namely:
\begin{equation*}
 \sum_{\mathcal{P}} A_\mathcal{P}(k) = 1, \quad \forall k \in \{0, 1, \ldots, 6\}.   
\end{equation*}

\noindent As a result, only three out of the four phase coefficients are independent for each value of $k$, and the coefficient for the cubic phase $A_C(k)$ is fully determined by those of R, O, and T. The model therefore contains $3 \times 7 = 21$ independent parameters.
Additional physical constraints can be introduced to further reduce this number. In particular, we impose that for $x$ = 0 (BTO), only the rhombohedral phase is observed, and for $x$ = 1 (BZO), only the cubic phase is present. This leads to the conditions:
$A_R(6) = 1, A_{O}(6) = A_T(6) = A_C(6) = 0, A_C(0) = 1, A_R(0) = A_{O}(0) = A_T(0) = 0.$
It is worth noting that the condition corresponding to $x$=0, a Ti site surrounded by six Ti neighbors, can also occur at intermediate compositions. Similarly, the condition for $x$=1, a Ti site with zero Ti neighbors, may appear for other $x$ values.
In fact, this latter situation naturally arises in a Ti/Zr rock-salt type arrangement, in which Ti ions become non-polar due to the absence of polar Ti neighbors, a result obtained both from first-principles calculations\cite{amor_18} and reproduced by the present model.
Moreover, we assume a sequential phase evolution: in the low-concentration regime ($x \approx 0$), Ti cells may transform from R to O as Zr is introduced; conversely, in the high-concentration regime ($x \approx 1$), Ti cells may evolve from C to T as Ti neighbors are added. These assumptions can be encoded as $A_R(5) = A_{O}(5) = 0.5,\quad A_T(5) = A_C(5) = 0, \quad A_T(1) = A_C(1) = 0.5,\quad A_R(1) = A_{O}(1) = 0$. 
With these constraints, the number of independent parameters is reduced to 9. These remaining parameters can be optimized by minimizing the mean absolute error between the model predictions and the numerical data.

\begin{figure}
\centering
\includegraphics[scale=0.6]{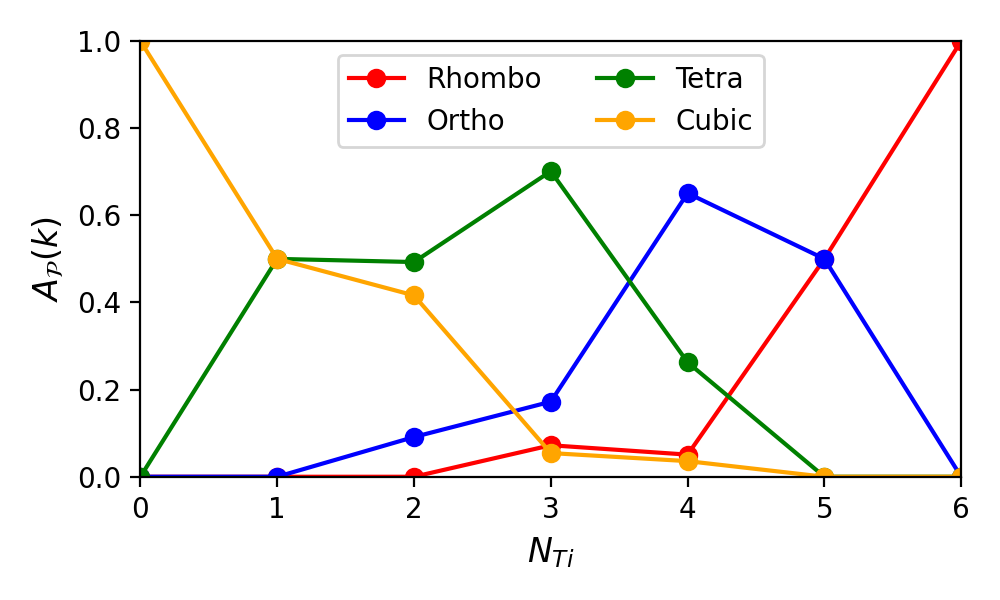}
\caption{\label{fig:Coef_x} Probability for a Ti-centered cell to adopt a specific polar phase as a function of the number of neighboring Ti cells in BaZr$_x$Ti$_{1-x}$O$_3$ with a random B-site distribution at $T = 0$ K. }
\end{figure}

As expected, the values obtained for the coefficients $A_\mathcal{P}(k)$ (Fig.\ref{fig:Coef_x}) confirm that the probability for a given cell to adopt a specific polar phase is directly determined by the number of neighboring Ti ions. 
Sites with more Ti neighbors tend to stabilize lower-symmetry polar phases, whereas those with fewer active neighbors favor higher-symmetry configurations such as T or C. For instance, the probability of finding an R cell with fewer than four Ti neighbors is negligible. This simple probabilistic framework consistently accounts for the observed phase fractions and their evolution with composition.
It is important to note, however, that this simplified model neglects several potentially relevant factors. For example, Zr cells are treated purely as inactive, without accounting for their possible influence on nearby Ti cells or on the surrounding lattice through effects such as local strain. Moreover, the model ignores the spatial arrangement and directional distribution of Ti neighbors, both of which can play a significant role in local symmetry breaking and phase stability.
It should also be noted that an overestimation of the polarization threshold would mainly result in an overestimation of the fraction of the cubic (non-polar) phase, while the relative proportions of the polar phases (T, O, and R) would remain largely unaffected.
Despite these simplifications, the model successfully captures the main trends in phase distribution and serves as a useful approximation for understanding the composition-dependent behavior in BZT with a random B-site distribution.

\subsection{\label{sec:phdl} Finite temperature properties} 

\subsubsection{Phase diagram}
Using data obtained at zero temperature and molecular dynamics simulations, we constructed the temperature–composition phase diagram. For compositions up to $x$=0.3, where the ground state is ferroelectric, phase transitions were identified based on the behavior of polarization and dielectric permittivity as a function of temperature. At higher Zr concentrations, the system does not exhibit a long-range ferroelectric order,  and we adopt as an order parameter T$_m$, the point where the dielectric susceptibility reaches its maximum value. In this case, we calculated the susceptibility using the direct method, which involves analyzing changes in polarization as a function of an applied electric field.

\begin{figure}
\centering
\includegraphics[width=\linewidth]{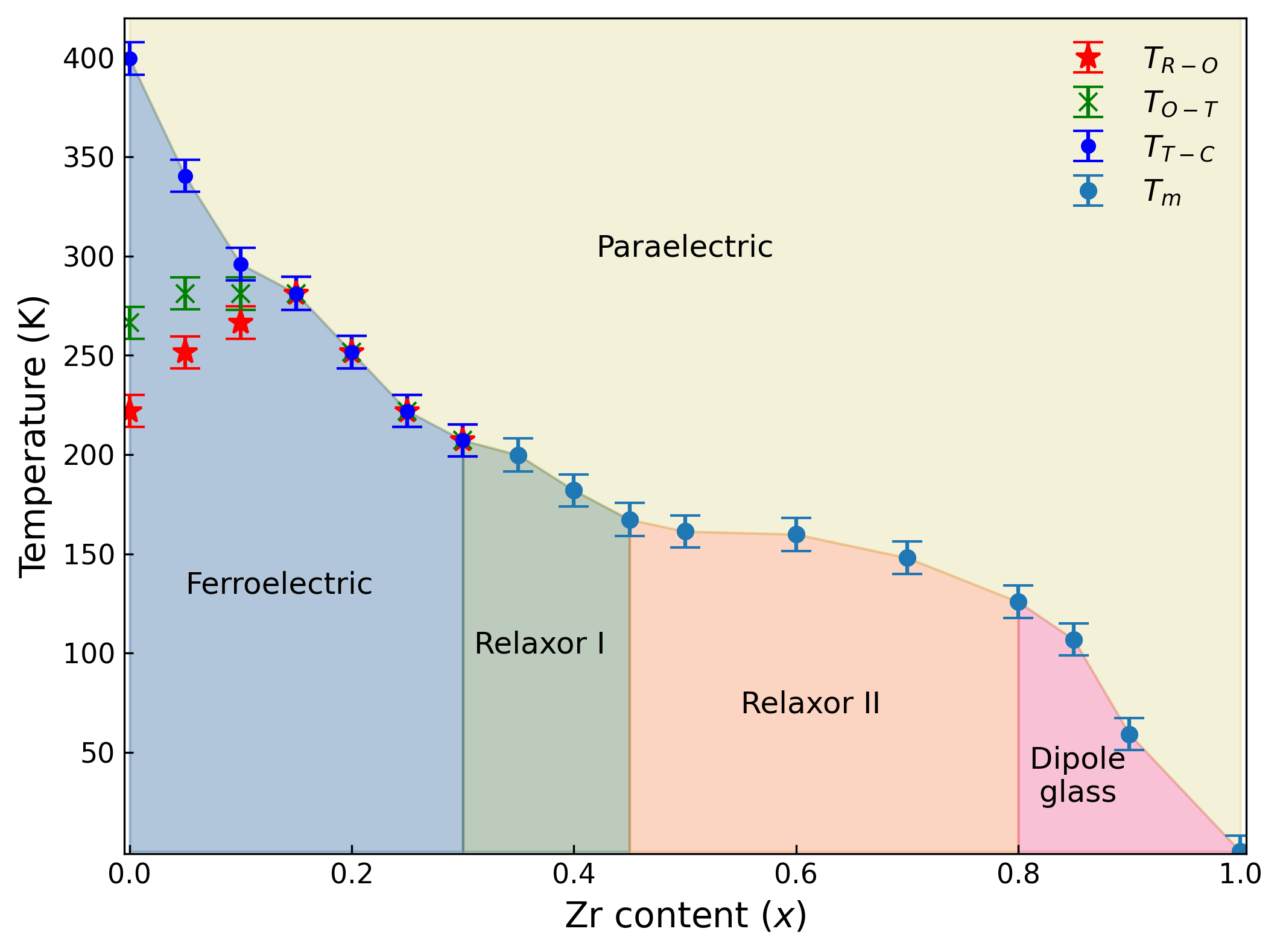}
\caption{\label{fig:PhD}Phase diagram of BZT as a function of Zr content obtained from MD simulation with the atomistic model.  T$_{R-O}$, T$_{O-T}$, T$_{T—C}$ denote the transitions between the rhombohedral (R),
orthorhombic (O), tetragonal (T), and cubic (C) phases. T$_m$ represents the temperature corresponding to the maximum in dielectric permittivity. }
\end{figure}

The resulting phase diagram obtained  with the model (Fig. \ref{fig:PhD} ) is in excellent agreement with experiments\cite{dong_12,shva_2012,maiti_11,petz_21} and previous simulations\cite{ment_19,mayer_22}. 
The model, constructed based on the properties of the end-member compounds, effectively captures the compositional and thermal evolution of the solid solution. At low Zr concentrations, the simulated BZT system exhibits the same sequence of ferroelectric phases as pure BTO. As $x$ increases, the transition temperatures for the R–O and O–T phases shift to higher values, whereas the T–C transition shifts downward. These transition lines converge near $x=0.15$, beyond which the distinction between the intermediate O and T phases becomes increasingly diffuse. In the range $0.15<x<0.30$, a single R ferroelectric phase remains stable, with T$_C$ steadily decreasing as Zr content increases. At higher Zr concentrations, a macroscopically non-polar state is stabilized at all temperatures, signaling the absence of long-range ferroelectric order.  Within this regime, three compositional subranges can be identified, consistent with experimental observations \cite{petz_21}. These include two types of relaxor regions and a dipolar glass state. Specifically, the relaxor type I region, for $0.30 < x < 0.45$, is characterized by a temperature T$_m$ that decreases with increasing x. The relaxor type II region, for $0.45 < x < 0.8$, shows an approximately temperature-independent T$_m$. For $x > 0.8$, the system enters a dipolar glass state, where the characteristic temperature gradually decreases and ultimately vanishes at $x$=1.

\begin{figure*}
\centering
\includegraphics[width=\linewidth]{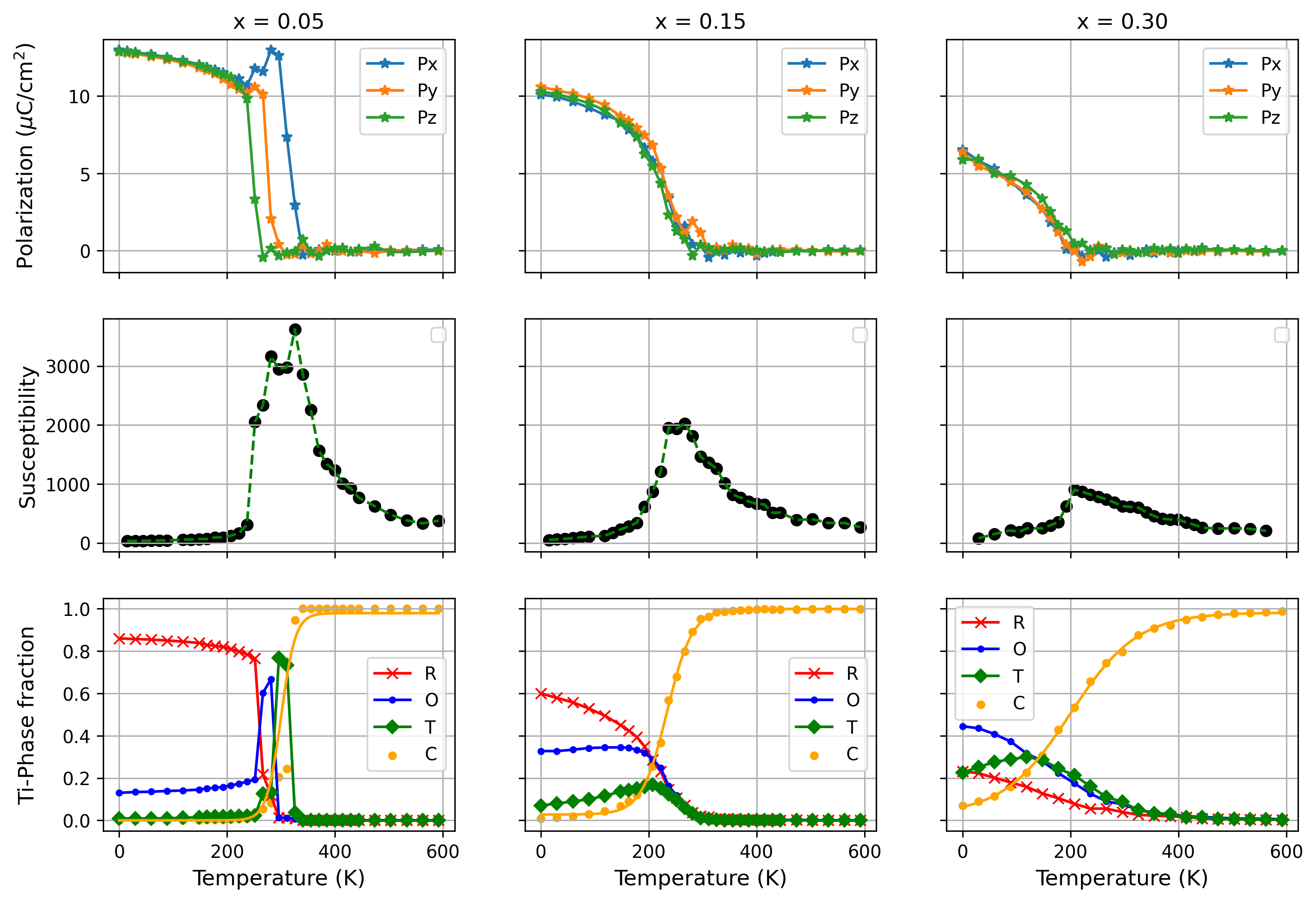}
\caption{\label{fig:Ferr_v_T_new}Temperature dependence of polarization (top row), dielectric susceptibility (middle row), and the fraction of Ti-centered cells in each structural phase (bottom row) for BZT compositions x=0.05, 0.15, and 0.30. }
\end{figure*}
To explore how local structural and polar features give rise to the macroscopic phase diagram, the analysis of local phases in Ti-centered cells can be naturally extended to finite temperatures. This approach provides a detailed and consistent microscopic understanding of the phase transitions.  The results demonstrate that phase coexistence is a key feature of BZT, extending across both composition and temperature ranges and highlighting the complex interplay between local structural order and macroscopic ferroelectric behavior.

\subsubsection{Ferroelectric regime}
Fig.  \ref{fig:Ferr_v_T_new}  presents results for three representative compositions within the ferroelectric regime ($x = 0.05, 0.15$, and $0.30$). The top panels display the temperature dependence of the macroscopic polarization, while the bottom panels show the corresponding evolution of the local phase distribution in Ti cells. In addition, the middle panels show the dielectric susceptibility, which is also computed at higher Zr concentrations,  where macroscopic polarization is no longer a suitable order parameter, thus enabling a more consistent comparison across the full compositional range, both with simulation results and with experimental observations. 
 
 At $x = 0.05$, BZT exhibits a behavior closely resembling that of pure BTO, with sharp and well-defined phase transitions characterized by abrupt changes in the components of the macroscopic polarization, as well as distinct peaks in the dielectric susceptibility.
As shown in the figure, these transitions can also be tracked through the evolution of local symmetries in Ti cells, which follow the macroscopic sequence R–O–T–C. Although some cells exhibit a local symmetry different from the macroscopic phase—for instance, O-type cells within the R phase, T-type cells emerging during the O phase, or residual R-type cells persisting above the R–O transition—these deviations are relatively minor and do not alter the dominant phase character. Non-polar C-type cells begin to appear during the O phase, gradually increasing with temperature and fully dominating the system above T$_C$.

At $x = 0.15$,  the three phase transitions characteristic of BTO become merged or pinched into a direct R-C phase transition~\cite{maiti_08,maiti_11} . Macroscopically, the three polarization components are similar in magnitude and  decrease continuously with increasing temperature up to T$_C$, accompanied by a single broad peak in the dielectric susceptibility that reproduces the experimentally observed behavior at this composition\cite{hen_1982}. Microscopically, however, the system undergoes  a more intricate evolution, evident in the gradual redistribution of local phase symmetries. Although R-type cells remain the most prevalent at low temperatures, the increased Zr concentration leads to a noticeable rise in the number of O- and T-type cells. 
As the temperature rises, the R fraction steadily decreases. Simultaneously, the O and T fractions show a slight increase, peaking at approximately 180 K and 220 K, respectively, just before the phase transition. These inflection points, while not representing an abrupt phase transition, mark the onset of a redistribution of local symmetries. It may be interpreted as the points where the stability of the O and T local environment becomes compromised due to thermal fluctuations.  
This variation in local phase populations reflects a sequence of continuous local transformations. The decline in R cells corresponds to their gradual conversion into O and, eventually, T and C types. O-type cells initially grow in number as they emerge from R regions but are later depleted as they transition into T or C states. Similarly, T-type cells arise primarily from O (and to a lesser extent R) cells, yet eventually become nonpolar as well. This cascade of local reconfigurations underlies the continuous redistribution of phase types, which, in turn, mirrors the macroscopic suppression of ferroelectric order and the diffuse nature of the transition.

At $x = 0.3$, the increased Zr content leads to a reduction in macroscopic polarization, and the R–C transition becomes more gradual. In this case, the dielectric susceptibility displays a broader, less intense peak shifted to lower temperatures, reflecting the increased diffuseness of the transition.
At this composition, O cells become more abundant than R ones, and the fractions of both polar phases progressively decline with increasing temperature. Concurrently, the proportion of T cells rises, reaching a maximum near 150 K. Above this temperature, all polar cell types gradually diminish until they disappear. Notably, some polar cells persist even above T$_C$ before eventually disappearing, indicating a gradual transition toward the non-polar phase. 

This gradual redistribution of local phase types is consistent with experimental observations of diffuse phase transitions in BZT. Dielectric measurements at similar compositions reveal broadened peaks and frequency-dependent responses, which are typically associated with the coexistence of multiple local symmetries and the lack of a sharp transition. The persistence of polar cells above T$_C$, along with the overlapping presence of R, O, and T types, mirrors the experimental evidence of local polarization in the nominally paraelectric state. These findings support the view that the transition is not abrupt, but rather the result of continuous, temperature-driven reorganization of local dipolar environments—a hallmark of relaxor ferroelectric behavior.

\subsubsection{Non-ferroelectric regime}
\begin{figure*}
\centering
\includegraphics[width=\linewidth]{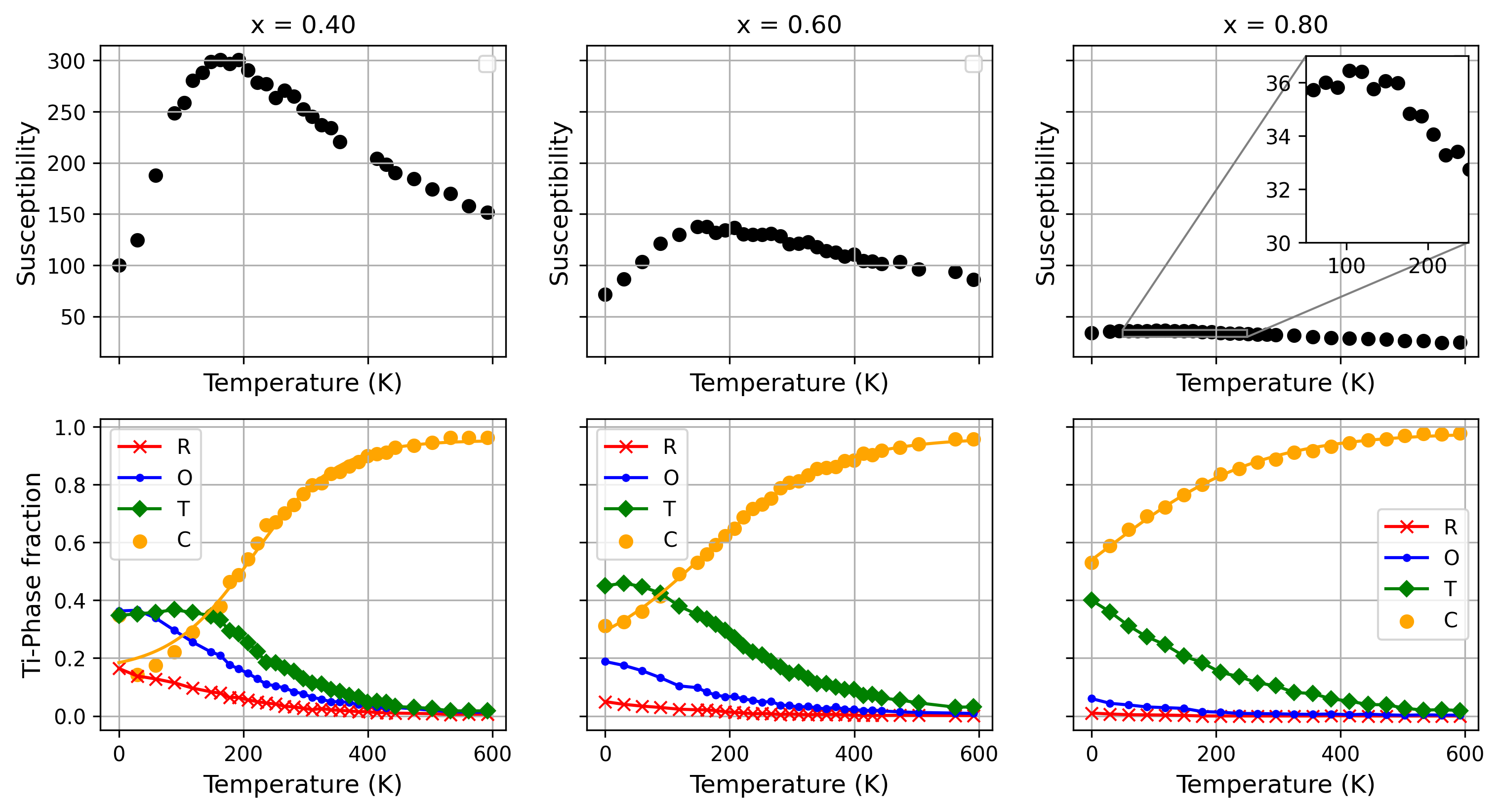}
\caption{\label{fig:Ferr_v_T2}Temperature dependence of the dielectric susceptibility (top row) and the fraction of Ti-centered cells in each structural phase (bottom row) for BZT compositions $x$=0.40, 0.60, and 0.80. }
\end{figure*}

At higher Zr concentrations, macroscopic polarization is negligible, and the analysis focuses on the dielectric susceptibility and the evolution of local phase fractions. Fig.~\ref{fig:Ferr_v_T2} presents the temperature dependence of these quantities for representatives $x$ in this region. $x$ = 0.4, 0.6, and 0.8.
For the dielectric susceptibility (top panel), it continues the trend observed at lower Zr content, with broader and less intense peaks as the amount of Zr increases. The peak positions also shift to lower temperatures with increasing $x$, occurring at approximately 192 K, 162 K, and 100 K for $x=$ 0.4, 0.6, and 0.8, respectively. In contrast to the lower-$x$ region, here the fractions of the different types of local cells vary smoothly with temperature. The polar behavior is primarily governed by Ti-centered cells with local T symmetry, with minor contributions from R and O cells. A deeper analysis of the nonpolar C cell population reveals the existence of two distinct regimes. For $x=$0.4 and 0.6, both curves display an inflection point with temperature, marking a characteristic change in the rate of transformation. This evolution is well captured by a function of the form:

\begin{equation}
f(T)=\frac{1}{2}\left [1+tanh \left (\frac{T - T_0}{\Delta T} \right) \right]
\label{tanh}
\end{equation}

\noindent which smoothly evolves with increasing temperature. In the expression, $T_0$ marks the inflection point of the continuous temperature-induced transformation, while $\Delta T$ determines the broadness of the variation. Notably, the fit yields $T_0$ = 192 K for $x=$0.4 and 150 K for $x=$0.6,  which closely coincide with the temperatures $T_m$ previously obtained from the susceptibility. In other words, the maximum in dielectric susceptibility corresponds to the point where the rate of increase (change) in the fraction of nonpolar cells is highest. This agreement highlights the connection between the local structural evolution and the dielectric response of the system.
In contrast, at $x=$0.8, the evolution of T and C fraction cells is monotonic, and no inflection point can be clearly identified. As a result, the correspondence between $T_0$ and $T_m$ is lost, suggesting that a different mechanism may be governing the dielectric response at this composition. This behavior may indicate a transition from relaxor-like dynamics to a dipolar glass regime, in which structural and dielectric features become increasingly decoupled. 

While the current framework successfully explains the BZT phase diagram through local-phase distributions analysis, the present study does not address the role of spatial correlations of local polarizations, which is of fundamental interest for understanding relaxor behavior. Future work will therefore focus on investigating the formation of polar nanoregions, or "slush-like" polar states, characterized by multi-domains with extremely small sizes \cite{yuan_23,hana_25}. A deeper understanding of these structures is essential to provide more detailed insight into the microscopic origins of relaxor ferroelectricity in BZT solid solutions.

\section{\label{sec:conclu} Conclusions}

In summary, we have developed a first-principles–based atomistic framework to describe the complex phase diagram of BZT. By classifying individual unit cells according to their local symmetry, we showed that the macroscopic behavior of BZT can be understood in terms of the coexistence of symmetry-distinct polar phases. The relative fractions of R, O, T, and C phases evolve smoothly with composition and temperature, and their stability appears to be strongly influenced by the local chemical environment, particularly the number of neighboring Ti ions. This probabilistic picture provides a consistent explanation for the progressive replacement of long-range ferroelectric order by nanoscale phase coexistence as Zr content increases, accounting for the crossover from sharp ferroelectric transitions to the diffuse features of relaxor and dipolar-glass states. Overall, our results highlight the critical role of local atomic disorder in governing the functional properties of complex perovskite solid solutions. Beyond the present results, the phase-coexistence framework introduced here opens a way to a deeper exploration of the relaxor behavior of BZT and related systems.

\begin{acknowledgments}
This work was sponsored by the National
Scientific and Technical Research Council (Consejo Nacional de Investigaciones Científicas y Técnicas, CONICET) and the National University of Rosario (Universidad Nacional de Rosario, UNR) (No.~80020180300068UR).
The results presented in this work have been obtained by using the facilities of the CCT-Rosario Computational Center, a member of the High Performance Computing National System (SNCAD, MincyT-Argentina). FDR acknowledges the financial support of the European Union by the ERC-STG project 2D-sandwich (Grant No 101040057).
\end{acknowledgments}

\bibliography{bzt_biblio}


\end{document}